\documentclass[a4paper]{article}

\usepackage{INTERSPEECH2020}
\usepackage{CJKutf8}

\usepackage{threeparttable}
\usepackage{multirow}
\usepackage{booktabs}
\usepackage{amsmath}

\usepackage{soul}

\newcommand{\eg}{e.\,g.,\ }
\newcommand{\ie}{i.\,e.,\ }

\title{An Early Study on Intelligent Analysis of Speech under COVID-19: \\Severity, Sleep Quality, Fatigue, and Anxiety}

\name{Jing Han$^1$, Kun Qian$^{2*}$,  Meishu Song$^1$, Zijiang Yang$^1$, Zhao Ren$^1$, Shuo Liu$^1$,\\Juan Liu$^{3*}$, Huaiyuan Zheng$^{4*}$, Wei Ji$^{4*}$, Tomoya Koike$^2$, Xiao Li$^5$, Zixing Zhang$^6$,\\Yoshiharu Yamamoto$^2$,  and Bj\"orn W.\ Schuller$^{1,6}$}
\address{ \small 
  $^1$ Chair of Embedded Intelligence for Health Care and Wellbeing, University of Augsburg, Germany\\
  $^2$ Educational Physiology Laboratory, University of Tokyo, Japan\\
  $^3$ Department of Plastic Surgery, Central Hospital of Wuhan, Tongji Medical College, \\ Huazhong University of Science and Technology, China \\
  $^4$ Department of Hand Surgery, Wuhan Union Hospital, Tongji Medical College, \\ Huazhong University of Science and Technology, China\\
  $^5$ Department of Neurology, Children’s Hospital of Chongqing Medical University, China\\
  $^6$ GLAM -- Group on Language, Audio \& Music, Imperial College London, UK\\
  \thanks{$^{*}$Kun Qian, Juan Liu, Huaiyuan Zheng, and Wei Ji are the \emph{Corresponding Authors}. E-mails:~qian@p.u-tokyo.ac.jp, \{liujuan\_1018,~zhenghuaiyuan\}@126.com, jiwei1230@foxmail.com}.}
\email{jing.han@informatik.uni-augsburg.de}

\begin{document}

\maketitle
\begin{abstract}
The COVID-19 outbreak was announced as a global pandemic by the World Health Organisation in March 2020 and has affected a growing number of people in the past few weeks. In this context, advanced artificial intelligence techniques are brought 
to the fore 
in responding to fight against and reduce the impact of this global health crisis. In this study, we focus on developing some potential use-cases of intelligent speech analysis for COVID-19 diagnosed patients. In particular, by analysing speech recordings from these patients, we construct audio-only-based models to automatically categorise the health state of patients from four aspects, including the severity of illness, sleep quality, fatigue, and anxiety. For this purpose, two established acoustic feature sets and support vector machines are utilised. Our experiments show that an average accuracy of $.69$ obtained estimating the severity of illness, which is derived from the number of days in hospitalisation. We hope that this study can foster an extremely fast, low-cost, and convenient way to automatically detect the COVID-19 disease. 


\end{abstract}
\noindent\textbf{Index Terms}: COVID-19 diagnosis, speech analysis, computational paralinguistics

\section{Introduction}

The emergence and spread of the novel coronavirus 
and the related COVID-19 disease is deemed as a major public health threat for almost all countries around the world. Moreover, explicitly or implicitly, the coronavirus pandemic has brought an unprecedented impact on people across the world. To combat the COVID-19 pandemic and its consequences, clinicians, nurses, and other care providers are battling in the front-line. Apart from that, scientists and researchers from a bench of research domains are also stepping up in response to the challenges raised by this pandemic. For instance, several different kinds of drugs and vaccines are being developed and trialled, to treat the virus or to protect against it~\cite{Cynthia20-RAD, ren2020traditional, cao2020covid, Peeples20-NRA, Nicole20-DCV}, and meanwhile methods and technologies are designed and investigated to accelerate the diagnostic testing speed\cite{DURNER2020, Chio20-IST}.


Particularly, considering the community of data science, massive efforts have been and are still being made to mine information data-driven. In particular, 
a number of works have been proposed to promote automatic screening by analysing chest CT images~\cite{li2020artificial, afshar2020covid, wang2020covid, farooq2020covid}. For instance, in~\cite{li2020artificial}, the deep model COVNet was developed to extract visual features to detect COVID-19. However, no research work has yet been done to explore sound-based COVID-19 assessment.


In the perspective of sound analysis, as coronavirus is a respiratory illness, abnormal breathing patterns from patients intuitively might be a potential indicator for diagnosis~\cite{wang2020abnormal}. 
According to the latest clinical research, the severity of the COVID-19 disease
can be categorised into three levels, namely, mild, moderate, and severe illness~\cite{cascella2020features}. For each level, various typical respiratory symptoms can be observed, from dry cough presented in mild illness, to shortness of breath in moderate illness, and further to severe dyspnea, respiratory distress, or tachypnea (respiratory frequency $>$ 30 breaths/min) in severe illness~\cite{cascella2020features}. Meanwhile, all these breathing disorders lead to abnormal variations of articulation. Consequently, it can be of great interest to use automatic speech and voice analysis to aid COVID-19 diagnosis, which is non-invasive and low-cost.

In addition, there could be many meaningful and powerful audio-based tools and applications, which are so far underestimated, and hence underutilised. Pretty recently, scientists elaborated several potential use-cases in the fight against COVID-19 spread via exploiting intelligent speech and sound analysis~\cite{schuller2020covid}. Specifically, these envisioned directions are grouped into three categories, \ie audio-based risk assessment, audio-based diagnosis, and audio-based monitors such as for monitoring of spread, social distancing, treatment and recovery, and patient wellbeing~\cite{schuller2020covid}.

Albeit the importance of the work by analysing voice or speech signals to battle this virus pandemic, no empirical research work has been done to date. To fill this gap, we present an early study on the intelligent analysis of speech under COVID-19. To the best of our knowledge, this is the first work towards this direction. Particularly, we take a data-driven approach to automatically detect the patients' symptom severity, as well as their physical and mental states. It is our hope that this step can help develop a rapid, cheap, and easy way to diagnose the COVID-19 disease, and assist medical doctors.



\section{Data Collection} 
\label{sec:data}
For this work, getting collected and annotated data is the first step. At present, data collection is underway worldwide from both infected patients at various stages of the disease, and healthy individuals as a control group. For instance, the App ``COVID-19 Sounds App''\footnote{https://www.covid-19-sounds.org/en/} has been released from the University of Cambridge to help researchers detect if a person is suffering from COVID-19, by collecting recordings of participants' voice, their breathing, as well as coughing. 
Likewise, researchers from Carnegie Mellon University launched a new App ``Corona Voice Detect''\footnote{https://cvd.lti.cmu.edu/} to gather voice samples, such as coughs, several vowel sounds, counting up to twenty, and the alphabet. Nonetheless, currently all these data are not publicly available for research purposes.

\begin{figure}[!t]
    \centering
    \includegraphics[width=0.499\textwidth, clip]{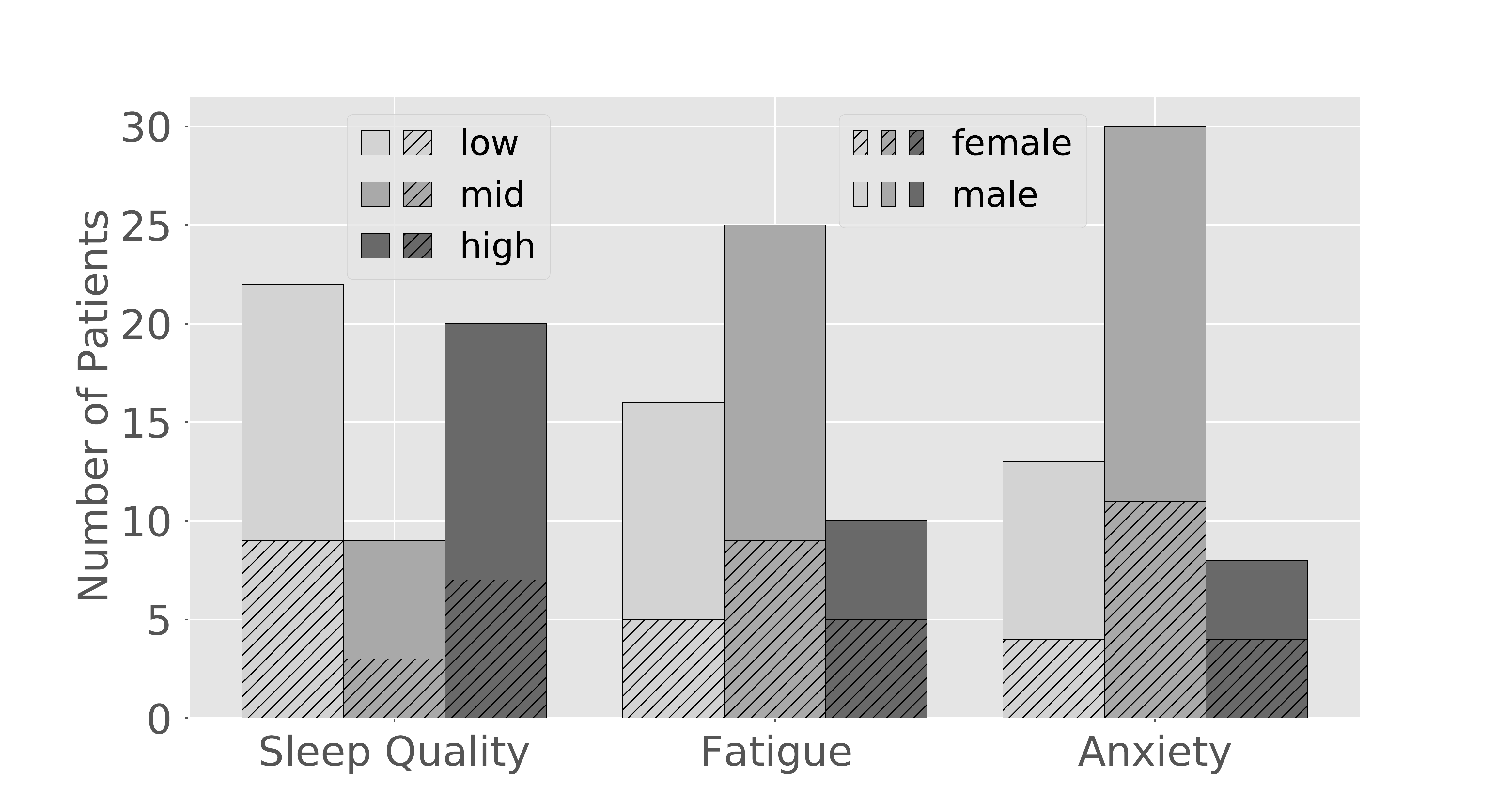}
    \caption{Distribution of 51 COVID-19 patients' self-reported questionnaires regarding to their health states in three categories, namely, sleep quality, fatigue, and anxiety. For each category, the patient is asked to select one of  three degrees (\ie low, mid, and high).}
    \label{fig:my_label}
      \vspace{-.3cm}

\end{figure}

We collected in-the-wild individual-case data of 52 COVID-19 infected hospitalised patients (20 females and 32 males) from two hospitals in Wuhan, China. Data were collected between 20 to 26 March 2020. For each patient, five sentences were recorded one after another via the Wechat App, when doctors and nurses were making their daily rounding to check the patients. As a result, five recordings per patient were acquired during data collection. Note that, these five sentences all have neutral meaning and one example (relating to the date and days as variables) is  given below:
\begin{CJK*}{UTF8}{gbsn}
\begin{enumerate}
    \item 今天是2020年3月26号。
    \item 我同意使用我的语音进行与肺炎相关的研究。
    \item 这是我住院的第12天。
    \item 我很想早点康复出院。
    \item 今天的天气是晴天。
\end{enumerate}
\end{CJK*}
(when translated into English)
\begin{enumerate}
    \item Today is 2020 March 26th.
    \item I agree to use my voice for coronavirus-related research purposes.
    \item Today is the twelfth day since I stayed in the hospital.
    \item I wish I could rehabilitate and leave hospital soon.
    \item The weather today is sunny.
\end{enumerate}
Moreover, three self-report questions were answered by each patient, regarding her (or his) sleep quality, fatigue, and anxiety. Specifically, participants rated their sleep quality/ fatigue/ anxiety by choosing from three different levels (\ie low, mid, and high). Furthermore, regarding demographic information, another four characteristics of the patients were collected, including age, gender, height, and weight. Note that, the height and weight information from 13 patients were not provided. A statistical overview of the data can be seen in Table~\ref{tab:covid}.
\begin{table}[!t]
\centering
 \caption{Statistics of the COVID-19 audio data. }
 \begin{threeparttable}
  \begin {tabular}{lccc}
  \toprule
   {mean$\pm$std dev} &   female &   male &   all genders \\
  \midrule
age [years]  &  64.4{$\pm$9.5}   & 62.7$\pm$9.6 & 63.4$\pm$9.9 \\
height [cm] &  158.8$\pm$9.0 &  173.4$\pm$9.2 & 167.2$\pm$9.2  \\
weight [kg] &  59.1$\pm$13.6 & 71.0$\pm$14.1 &  66.0$\pm$13.5 \\
  \bottomrule
  \end{tabular}
    \vspace{-.3cm}
 \end{threeparttable}
 \label{tab:covid}
\end{table}
Furthermore, a distribution of the self-reported sleep quality, fatigue, and anxiety, grouped by gender, is illustrated in Fig.~\ref{fig:my_label}.

\section{Data Preprocessing} 
Once the COVID-19 audio data were collected, a series of data preprocessing processes were implemented. Specifically, we did the following four processes: data cleansing, hand-annotating of voice activities, speaker diarisation, and speech transcription. Details are provided below.

\noindent\textbf{Data Cleansing}: as the recordings were collected in the wild, there were few unsuccessful recordings where the patient failed to provide any speech rather than noisy background. In such cases, the recordings were discarded for further analysis. As a result, recordings from one female patient were discarded, resulting in data from only 51 subjects for further processing.

\noindent\textbf{Voice Activity Detection}: for each recording, then, the presence or absence of human voice was detected manually in Audacity\footnote{https://www.audacityteam.org/}. This is because, for some recordings, there were silence periods for the first few seconds and/or the last few seconds. Note that, we only removed the beginning and the end of unvoiced parts from each recording where no audible breathing took place. Hence, only voiced segments (\eg speech, breathing, and coughing) from the recordings were maintained. 

\noindent\textbf{Speaker Diarisation}: among the remaining voiced segments, there is speech from other individuals than the targeted patient. For this reason, we manually checked and annotated the speaker identities for each voiced segment, indicating if the voice was generated by the patient, or by anyone else.

\noindent\textbf{Speech transcription}: after annotating the speaker identities, voiced segments from the targeted diagnosed patients were converted into text transcriptions. Note that, while the collection was in the wild, beyond the aforementioned five sentences, some spontaneous recordings were spoken by the patients but with impromptu and unscripted content. 

After data preprocessing, we obtained in total of 378 segments. For this preliminary study, we focus only on the scripted segments from patients, leading to 260 pieces of recordings for further analysis. A statistic of the distribution of the five sentences is provided in Table~\ref{tab:5sent}. It can be seen that the distribution is imbalanced. The reason is that, some patients recorded the same content more than once, while some patients failed to supply all five recordings.
\begin{table}[!t]
\centering
 \caption{Distribution of the five targeted sentences collected from 51 COVID-19 diagnosed patients.}
 \begin{threeparttable}
  \begin {tabular}{ccccccc}
  \toprule
  sentence &   1 &  2 &   3 & 4 & 5  & $\sum$\\
  \midrule
\#   &  53   & 47 & 60 & 51 & 49 & 260\\
  \bottomrule
  \end{tabular}
  \vspace{-.3cm}
 \end{threeparttable}
 \label{tab:5sent}
\end{table}

These 260 audio segments from 51 COVID-19 infected patients, were then converted to mono signals with a sampling rate of 16\,kHz for further analysis. 

\section{Experiments and Results}
In this section, we detail out the experiments that were exectued to verify the feasibility of audio-only-based COVID-19 patient state assessment. More specifically, we first describe the experimental setups including the applied acoustic feature sets as well as related evaluation strategies. Afterwards, we elaborate on the experiment performance for COVID-19 severity estimation, as well as prediction performance of three COVID-19 patient self-reported status attributes, namely, sleep quality, fatigue, and anxiety degrees. Last, we discuss the limitation of the current study, and provide future work plans and directions.

\subsection{Feature Extraction}
Two established acoustic feature sets were considered in this study, namely, the Computational Paralinguistics Challenge (\textsc{ComParE}) set and the extended Geneva Minimalistic Acoustic Parameter Set ({eGeMAPS}). Specifically, these feature sets were extracted with the openSMILE toolkit~\cite{Eyben10-openSMILE}.

The \textsc{ComParE} feature set is a large-scale brute-force set utilised in the  series of INTERSPEECH Computational Paralinguistics Challenges since 2013~\cite{Schuller13-TI2, Schuller19-TI2}. It contains 6\,373 static features by computing various statistical functionals over 65 low-level descriptor (LLD) contours~\cite{Schuller13-TI2}. 
These LLDs consist of spectral (relative spectra auditory bands 1-26, spectral energy, spectral slope, spectral sharpness, spectral centroid, etc.), cepstral (Mel frequency cepstral coefficient 1-14), prosodic (loudness, root mean square energy, zero-crossing rate, $F_0$ via subharmonic summation, etc.), and voice quality features (probability of voicing, jitter, shimmer and harmonics-to-noise ratio). 
For more details, the reader is referred to~\cite{Schuller13-TI2}.

Different from the large-scale \textsc{ComParE} set, the other feature set applied in this work, eGeMAPS, is considerably smaller. It consists of only 88 features derived from 25 LLDs.
Particularly, these features were chosen concerning their capabilities to describe affective physiological changes in voice production. For more details about these features, please refer to~\cite{Eyben16-TGM}.

\subsection{Evaluation Strategy}
In this work, we carried out four audio-based classification tasks. First, we performed COVID-19 severity estimation based on the number of days of the hospitalisation. The hypothesis is that, a COVID-19 patient is generally very sick at the early stage of the hospitalisation, and then recovers step by step. As a consequence, the patients were approximately grouped into three categories, \ie the high-severity stage for the first 25 days, the mid-severity stage between 25 and 50 days, and the low-severity stage after 50 days.
Besides, further three classification tasks considered in this study are to predict the self-reported sleep quality, fatigue, and anxiety levels of COVID-19 patients, the potential of which has been spotted in~\cite{schuller2020covid}.

For these classification tasks, we implemented Support Vector Machines (SVMs) with a linear kernel function as the classifiers for all experiments, due to its widespread usage and appealing performance achieved in intelligent speech analysis~\cite{schmitt2016border, Han17-Prediction}. Specifically, a series of complexity constants $C$ were evaluated in [$10^{-7}$, $10^{-6}$ , $\cdots$, $10^{-1}$, $10^{0}$]. Further, to deal with the imbalanced data during training, a class weighting strategy was employed to automatically adjust the $C$ values in proportion to the importance of each class. The SVMs were implemented in Python based on the scikit-learn library.

Moreover, for all experiments in this study, Leave-One-Subject-Out (LOSO) cross-validation evaluations were carried out to satisfy the speaker independence  evaluation constraint. 
In this context, all the 260 instances were divided into 51 speaker-independent folds, with each fold containing only instances from one patient. With the LOSO evaluation scheme, one of the 51 folds was used as the test set and the other folds were put together to form a training set to train an SVM model. Then, this process was repeated 51 times until all folds were utilised as the test set. Note that, for each folder, an on-line standardisation was applied to the test set by using the means and variations of the respective training partition.

Then, the average performance was computed over the predictions of all instances. In this work, we utilise three most frequently-used measures, \ie Unweighted Average Recall (UAR), the overall accuracy (also known as Weighted Average Recall or WAR), and the F1 Score (also known as F-score or F-measure) that is the harmonic mean of precision and recall.

\subsection{Severity Estimation}
\begin{figure}[!t]
    \centering
    \includegraphics[width=0.465\textwidth, clip]{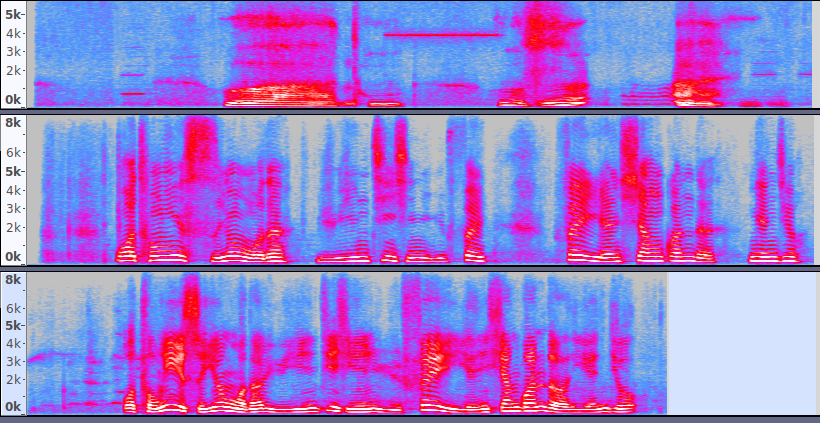}
    \caption{Illustration of the spectrograms from recordings by COVID-19 diagnosed patients with high (top), mid (middle), and low (bottom) severity degrees. All  were requested to speak the same content, \ie 
    the second sentence in the template (cf.~Section~\ref{sec:data}).}
    \label{fig:spec}
      \vspace{-.3cm}
\end{figure}

In Table~\ref{tab:res1}, we report the performance of the best SVM models for the two selected feature sets, respectively. In particular, the best model was chosen from varied SVMs with different $C$ values based on UAR. It can be seen that, the large feature set, ComPARE, performs slightly better than eGeMAPS for the severity estimation, achieving $.68$ UAR, $.69$ accuracy, and $.66$ F1-score.
Moreover, we further inspect the audio recordings from patients with varied severity  levels. An illustration is given in Fig.~\ref{fig:spec}. In particular, three recordings were taken from three different patients, who were asked to say the same content. The first patient failed to produce the sentence due to his severe symptoms. The second sample is from a female patient. She successfully spoke the whole content following the given template, however, had to pause several times to take a heavy breath before carrying on with the remaining content. In contrast, the third patient managed to generate the same recording more clearly and fluently.

\begin{table}[!t]
\centering
 \caption{Performance in terms of unweighted average recall (UAR), accuracy (WAR), and F1-score for three-class COVID-19 severity classification. The complexities $C$s that achieved the best performance are reported as well. Note: chance level is $.33$ for UAR.}
 \begin{threeparttable}
  \begin {tabular}{lcccc}
  \toprule
  \textbf{Feature} &  \textbf{$C$} & \textbf{UAR} &  \textbf{WAR} & \textbf{F1} \\
  \midrule
eGeMAPS   &  $10^{-6}$   & .67 & .69 & .65 \\
ComPARE   &   $10^{-6}$   & .68 & .69 & .66 \\
  \bottomrule
  \end{tabular}
    \vspace{-.3cm}
 \end{threeparttable}
 \label{tab:res1}
\end{table}



\subsection{Sleep Quality, Fatigue, and Anxiety Prediction}
Considering audio-based sleep quality, fatigue, and anxiety estimation, we further trained SVM models for each task separately. Corresponding results are shown in Table~\ref{tab:res2}, where the performance of the best models for each task and each feature set are provided. Similarly, the best performance was taken where the highest UAR was obtained, and performance in terms of accuracy and F1-score is given. 
When comparing the three tasks, the best performance is achieved for sleep quality classification, reaching up to $.61$ UAR. Then, for anxiety prediction, a UAR of $.56$ is attained. When it comes to fatigue prediction, the best performance of UAR is only $.46$, which is, however, above chance level ($.33$ for three-class classification).

Further, when comparing two selected feature sets, the compact eGeMAPS set consistently outperforms the large-scale ComPARE feature set. On the one hand, these results reveal the effectiveness of the eGeMAPS set for audio-based sleep quality, fatigue, and anxiety detection. On the other hand, the inferior performance based on ComPARE might be due to the low number of training samples.

\subsection{Discussion}
In this preliminary study, experiments were carried out based on speech recordings from COVID-19 infected and hospitalised patients. The results have demonstrated the feasibility and effectiveness of audio-only-based COVID-19 analysis, specifically in estimating the severity level of the disease, and in predicting the health and wellbeing status of patients including sleep quality, fatigue, and anxiety. Nonetheless, there are still many ways to extend the present study for further development. 

First, due to time limitation, the collected data set is relatively small, and lacks control group data from both healthy subjects and patients with other respiratory diseases. These data collections are still in progress for more comprehensive analysis in the future. In addition, AI techniques can be considered to tackle the data scarcity issue, such as data augmentation via generative adversarial networks~\cite{Goodfellow2014, Zhang19-Snore, Han19-Adversarial}. Given more data, the performance of our models is expected to be further improved and more robust.

Second, only functional features computed over whole segments were investigated. However, abnormal respiratory symptoms might be instantaneous and occur only in a short period of time. In this context, analysing low-level features in successive frames with sequential modelling might bring further performance improvement. Moreover, in addition to conventional handcrafted features, deep representation learning algorithms might be explored to learn representative and salient data-driven features for COVID-19 related tasks. These include deep latent representation learning~\cite{Han18-Emotion}, self-supervised learning~\cite{pascual2019learning}, and transfer learning~\cite{Ren18-LIR} to name but a few.

\begin{table}[!t]
\centering
 \caption{Performance in terms of unweighted average recall (UAR), accuracy (WAR), and F1-score for three three-class COVID-19 patient state tasks, \ie sleep (quality),  fatigue, and anxiety predictions. The SVM complexities $C$ that achieved the best performance are further given. Note: chance level is $.33$ for UAR.}
 \begin{threeparttable}
  \begin {tabular}{llcccc}
  \toprule
  \textbf{Task } & \textbf{Feature} &  \textbf{$C$} & \textbf{UAR} &  \textbf{WAR} & \textbf{F1} \\
  \midrule
sleep & eGeMAPS   &  $10^{-4}$   & .61 & .57 & .55 \\
sleep & ComPARE   &   $10^{-6}$   & .49 & .39 & .37 \\
  \midrule
fatigue & eGeMAPS   &  $10^{-6}$   & .46 & .50 & .42 \\
fatigue & ComPARE   &   $10^{-6}$   & .42 & .44 & .39 \\
  \midrule
anxiety & eGeMAPS   &  $10^{-3}$   & .56 & .50 & .49 \\
anxiety & ComPARE   &   $10^{-5}$   & .53 & .52 & .49 \\
  \bottomrule
  \end{tabular}
    \vspace{-.3cm}
 \end{threeparttable}
 \label{tab:res2}
\end{table}

Further, in this study, the severity estimation based on days in hospitalisation is in a rough fashion, as we are in lack of other information regarding the patients' health states. Similarly, the self-reported questionnaires about the patients' conditions are rather subjective, as they could have different principles and perception. If clinical examinations and reports of the patients are provided such as CT scans of their lungs~\cite{bernheim2020chest, pan2020time, zhao2020relation}, more objective and accurate labels could be attained.

Last but not least, in this paper, SVMs were separately trained to estimate four tasks. Considering the potential correlation between the severity of the disease, and the patient's sleep quality (or mood), a multi-task learning model might help effectively exploit the mutually dependent information from these tasks~\cite{zhang2016cross, parthasarathy2017jointly}.

\vspace{-.2cm}
\section{Conclusion}
At the time of writing this paper, the world has reported a total of 3\,020\,117 confirmed COVID-19 cases and 209\,799 fatalities, according to a dashboard developed and maintained by the Johns Hopkins University\footnote{https://coronavirus.jhu.edu/map.html}. 
To leverage the potential of computer audition to fight against this global health crisis, for the first time, experiments have been performed based on the speech of 51 COVID-19 infected and hospitalised patients from China. In particular, audio-based models have been constructed and assessed to predict the severity of the disease, as well as health and wellbeing-relevant mental states of the patients including sleep quality, fatigue, and anxiety. Experimental results have shown the great potential of exploiting audio analysis in the fight against COVID-19 spread.

In the future, we will continue the data collection process as well as collecting relevant clinical reports for a comprehensive understanding of the patient state. In addition, we attempt to introduce interpretable models and techniques to make the predictions more traceable, transparent, and trustworthy~\cite{8466590}. 

\vspace{-.2cm}
\section{Acknowledgements}
\footnotesize{
We express our deepest sorrow for those who left us due to COVID-19; they are lives, not numbers. We further express our highest gratitude and respect to the clinicians and scientists, and anyone else these days helping to fight against COVID-19, and at the same time help us maintain our daily lives. This work was partially supported by the Zhejiang Lab's International Talent Fund for Young Professionals (Project HANAMI), P.\,R.\,China, the JSPS Postdoctoral Fellowship for Research in Japan (ID No.\,P19081) from the Japan Society for the Promotion of Science (JSPS), Japan, and the Grants-in-Aid for Scientific Research (No.\,19F19081 and No.\,17H00878) from the Ministry of Education, Culture, Sports, Science and Technology (MEXT), Japan.}

\bibliographystyle{IEEEtran}

\bibliography{mybib}

\begin{thebibliography}{10}
\providecommand{\url}[1]{#1}
\csname url@samestyle\endcsname
\providecommand{\newblock}{\relax}
\providecommand{\bibinfo}[2]{#2}
\providecommand{\BIBentrySTDinterwordspacing}{\spaceskip=0pt\relax}
\providecommand{\BIBentryALTinterwordstretchfactor}{4}
\providecommand{\BIBentryALTinterwordspacing}{\spaceskip=\fontdimen2\font plus
\BIBentryALTinterwordstretchfactor\fontdimen3\font minus
  \fontdimen4\font\relax}
\providecommand{\BIBforeignlanguage}[2]{{%
\expandafter\ifx\csname l@#1\endcsname\relax
\typeout{** WARNING: IEEEtran.bst: No hyphenation pattern has been}%
\typeout{** loaded for the language `#1'. Using the pattern for}%
\typeout{** the default language instead.}%
\else
\language=\csname l@#1\endcsname
\fi
#2}}
\providecommand{\BIBdecl}{\relax}
\BIBdecl

\bibitem{Cynthia20-RAD}
C.~Liu, Q.~Zhou, Y.~Li, L.~V. Garner, S.~P. Watkins, L.~J. Carter, J.~Smoot,
  A.~C. Gregg, A.~D. Daniels, S.~Jervey, and D.~Albaiu, ``Research and
  development on therapeutic agents and vaccines for {COVID-19} and related
  human coronavirus diseases,'' \emph{ACS Central Science}, vol.~6, no.~3, pp.
  315--331, Mar. 2020.

\bibitem{ren2020traditional}
J.-l. Ren, A.-H. Zhang, and X.-J. Wang, ``Traditional chinese medicine for
  {COVID-19} treatment,'' \emph{Pharmacological research}, vol. 155, May 2020,
  2 pages.

\bibitem{cao2020covid}
X.~Cao, ``{COVID-19: I}mmunopathology and its implications for therapy,''
  \emph{Nature Reviews Immunology}, Apr. 2020, 2 pages.

\bibitem{Peeples20-NRA}
L.~Peeples, ``News feature: Avoiding pitfalls in the pursuit of a {COVID-19}
  vaccine,'' \emph{Proceedings of the National Academy of Sciences}, vol. 117,
  no.~15, pp. 8218--8221, Apr. 2020.

\bibitem{Nicole20-DCV}
N.~Lurie, M.~Saville, R.~Hatchett, and J.~Halton, ``Developing {Covid-19}
  vaccines at pandemic speed,'' \emph{New England Journal of Medicine}, Mar.
  2020, 5 pages.

\bibitem{DURNER2020}
J.~Durner, S.~Burggraf, L.~Czibere, T.~Fleige, A.~Madejska, D.~C. Watts,
  F.~Krieg-Schneider, and M.~Becker, ``Fast and simple high-throughput testing
  of {COVID 19},'' \emph{Dental Materials}, Apr. 2020, 4 pages.

\bibitem{Chio20-IST}
S.~Choi, C.~Han, J.~Lee, S.-I. Kim, and I.~B. Kim, ``Innovative screening tests
  for {COVID-19} in {South Korea},'' \emph{Clinical and Experimental Emergency
  Medicine}, Apr. 2020, 5 pages.

\bibitem{li2020artificial}
L.~Li, L.~Qin, Z.~Xu \emph{et~al.}, ``Artificial intelligence distinguishes
  {COVID-19} from community acquired pneumonia on chest {CT},''
  \emph{Radiology}, Mar. 2020, 16 pages.

\bibitem{afshar2020covid}
P.~Afshar, S.~Heidarian, F.~Naderkhani, A.~Oikonomou, K.~N. Plataniotis, and
  A.~Mohammadi, ``{COVID-CAPS: A} capsule network-based framework for
  identification of {COVID-19} cases from {X}-ray images,'' \emph{arXiv
  preprint arXiv:2004.02696}, 2020.

\bibitem{wang2020covid}
L.~Wang and A.~Wong, ``{COVID-Net: A} tailored deep convolutional neural
  network design for detection of {COVID-19} cases from chest radiography
  images,'' \emph{arXiv preprint arXiv:2003.09871}, 2020.

\bibitem{farooq2020covid}
M.~Farooq and A.~Hafeez, ``{Covid-resnet: A} deep learning framework for
  screening of {COVID19} from radiographs,'' \emph{arXiv preprint
  arXiv:2003.14395}, 2020.

\bibitem{wang2020abnormal}
Y.~Wang, M.~Hu, Q.~Li, X.-P. Zhang, G.~Zhai, and N.~Yao, ``Abnormal respiratory
  patterns classifier may contribute to large-scale screening of people
  infected with covid-19 in an accurate and unobtrusive manner,'' \emph{arXiv
  preprint arXiv:2002.05534}, 2020.

\bibitem{cascella2020features}
M.~Cascella, M.~Rajnik, A.~Cuomo, S.~C. Dulebohn, and R.~Di~Napoli, ``Features,
  evaluation and treatment coronavirus ({COVID-19}),'' in
  \emph{Statpearls}.\hskip 1em plus 0.5em minus 0.4em\relax StatPearls
  Publishing, 2020.

\bibitem{schuller2020covid}
B.~W. Schuller, D.~M. Schuller, K.~Qian, J.~Liu, H.~Zheng, and X.~Li,
  ``Covid-19 and computer audition: An overview on what speech \& sound
  analysis could contribute in the sars-cov-2 corona crisis,'' \emph{arXiv
  preprint arXiv:2003.11117}, 2020.

\bibitem{Eyben10-openSMILE}
F.~Eyben, M.~W\"ollmer, and B.~Schuller, ``{openSMILE} -- the {M}unich
  versatile and fast open-source audio feature extractor,'' in \emph{Proc.\ ACM
  International Conference on Multimedia (MM)}, Florence, Italy, 2010, pp.
  1459--1462.

\bibitem{Schuller13-TI2}
B.~Schuller, S.~Steidl, A.~Batliner \emph{et~al.}, ``The {INTERSPEECH} 2013
  computational paralinguistics challenge: social signals, conflict, emotion,
  autism,'' in \emph{{Proc.\ Annual Conference of the International Speech
  Communication Association (INTERSPEECH)}}, Lyon, France, 2013, pp. 148--152.

\bibitem{Schuller19-TI2}
B.~Schuller, A.~Batliner, C.~Bergler \emph{et~al.}, ``{The INTERSPEECH 2019
  Computational Paralinguistics Challenge: Styrian Dialects, Continuous
  Sleepiness, Baby Sounds \& Orca Activity},'' in \emph{{Proc.\ Annual
  Conference of the International Speech Communication Association
  (INTERSPEECH)}}, Graz, Austria, 2019, pp. 2378--2382.

\bibitem{Eyben16-TGM}
F.~Eyben, K.~Scherer, B.~Schuller, J.~Sundberg, E.~Andr{\'e}, C.~Busso,
  L.~Devillers, J.~Epps, P.~Laukka, S.~Narayanan, and K.~Truong, ``The geneva
  minimalistic acoustic parameter set {(GeMAPS)} for voice research and
  affective computing,'' \emph{IEEE Transactions on Affective Computing},
  vol.~7, no.~2, pp. 190--202, Apr. 2016.

\bibitem{schmitt2016border}
M.~Schmitt, F.~Ringeval, and B.~Schuller, ``At the border of acoustics and
  linguistics: Bag-of-audio-words for the recognition of emotions in speech,''
  in \emph{Proc.\ Annual Conference of the International Speech Communication
  Association (INTERSPEECH)}, San Francisco, CA, 2016, pp. 495--499.

\bibitem{Han17-Prediction}
J.~Han, Z.~Zhang, F.~Ringeval, and B.~Schuller, ``Prediction-based learning for
  continuous emotion recognition in speech,'' in \emph{Proc.\ IEEE
  International Conference on Acoustics, Speech and Signal Processing
  (ICASSP)}, New Orleans, LA, 2017, pp. 5005--5009.

\bibitem{Goodfellow2014}
I.~Goodfellow, J.~Pouget-Abadie, M.~Mirza, B.~Xu, D.~Warde-Farley, S.~Ozair,
  A.~Courville, and Y.~Bengio, ``Generative adversarial nets,'' in \emph{Proc.\
  the 27th Annual Conference on Neural Information Processing Systems (NIPS)},
  Montreal, Canada, 2014, pp. 2672--2680.

\bibitem{Zhang19-Snore}
Z.~Zhang, J.~Han, K.~Qian, C.~Janott, Y.~Guo, and B.~Schuller, ``{Snore-GANs:
  Improving Automatic Snore Sound Classification with Synthesized Data},''
  \emph{{IEEE Journal of Biomedical and Health Informatics}}, vol.~24, no.~1,
  pp. 300 -- 310, Jan. 2020.

\bibitem{Han19-Adversarial}
J.~Han, Z.~Zhang, N.~Cummins, and B.~Schuller, ``{Adversarial Training in
  Affective Computing and Sentiment Analysis: Recent Advances and
  Prospectives},'' \emph{{IEEE Computational Intelligence Magazine}, Special
  Issue on Computational Intelligence for Affective Computing and Sentiment
  Analysis}, vol.~14, no.~2, pp. 68--81, May 2019.

\bibitem{Han18-Emotion}
J.~Han, Z.~Zhang, G.~Keren, and B.~Schuller, ``Emotion recognition in speech
  with latent discriminative representations learning,'' \emph{Acta Acustica
  united with Acustica}, vol. 104, no.~5, pp. 737--740, Sep. 2018.

\bibitem{pascual2019learning}
S.~Pascual, M.~Ravanelli, J.~Serr{\`a}, A.~Bonafonte, and Y.~Bengio, ``Learning
  problem-agnostic speech representations from multiple self-supervised
  tasks,'' in \emph{{Proc.\ Annual Conference of the International Speech
  Communication Association (INTERSPEECH)}}, Graz, Austria, 2019, pp. 161--165.

\bibitem{Ren18-LIR}
Z.~Ren, N.~Cummins, V.~Pandit, J.~Han, K.~Qian, and B.~Schuller, ``{Learning
  Image-based Representations for Heart Sound Classification},'' in
  \emph{{Proc.\ 8th International Conference on Digital Health ({DH})}}, Lyon,
  France, 2018, pp. 143--147.

\bibitem{bernheim2020chest}
A.~Bernheim, X.~Mei, M.~Huang \emph{et~al.}, ``{Chest CT findings in
  coronavirus disease-19 (COVID-19): relationship to duration of infection},''
  \emph{Radiology}, Feb. 2020, 19 pages.

\bibitem{pan2020time}
F.~Pan, T.~Ye, P.~Sun \emph{et~al.}, ``{Time course of lung changes on chest CT
  during recovery from 2019 novel coronavirus (COVID-19) pneumonia},''
  \emph{Radiology}, Feb. 2020, 15 pages.

\bibitem{zhao2020relation}
W.~Zhao, Z.~Zhong, X.~Xie, Q.~Yu, and J.~Liu, ``Relation between chest {CT}
  findings and clinical conditions of coronavirus disease {(COVID-19)}
  pneumonia: a multicenter study,'' \emph{American Journal of Roentgenology},
  vol. 214, no.~5, pp. 1072--1077, May 2020.

\bibitem{zhang2016cross}
B.~Zhang, E.~M. Provost, and G.~Essl, ``Cross-corpus acoustic emotion
  recognition from singing and speaking: A multi-task learning approach,'' in
  \emph{Proc.\ IEEE International Conference on Acoustics, Speech and Signal
  Processing (ICASSP)}, Shanghai, China, 2016, pp. 5805--5809.

\bibitem{parthasarathy2017jointly}
S.~Parthasarathy and C.~Busso, ``Jointly predicting arousal, valence and
  dominance with multi-task learning,'' in \emph{{Proc.\ Annual Conference of
  the International Speech Communication Association (INTERSPEECH)}},
  Stockholm, Sweden, 2017, pp. 1103--1107.

\bibitem{8466590}
A.~{Adadi} and M.~{Berrada}, ``Peeking inside the black-box: A survey on
  explainable artificial intelligence {(XAI)},'' \emph{IEEE Access}, vol.~6,
  pp. 52\,138--52\,160, Sep. 2018.

\end{thebibliography}

\end{document}